\documentclass[%
 reprint,
 amsmath,amssymb,
 aps,
]{revtex4-2}

\usepackage{graphicx} 
\usepackage{amsmath,fullpage}
\usepackage{amssymb}
\usepackage{color}
\usepackage{mathtools}
\usepackage{dcolumn}
\usepackage{bm}

\def \be {\begin{equation}}
\def \ee {\end{equation}}
\def \bea {\begin{eqnarray}}
\def \eea {\end{eqnarray}}

\newcommand{\NN}{\ensuremath{\mathcal{N}}}
\newcommand{\cH}{\mathcal{H}}

\begin{document}

\title{A non-perturbative second law of black hole mechanics in effective field theory}
\author{Iain Davies}
\email{id318@cam.ac.uk}
\author{Harvey S. Reall}
\email{hsr1000@cam.ac.uk}
\affiliation{Department of Applied Mathematics and Theoretical Physics, University of Cambridge, Wilberforce Road, Cambridge CB3 0WA, United Kingdom}


\begin{abstract}
We describe a method for defining dynamical black hole entropy in gravitational effective field theories (EFTs). The entropy is constructed order by order in derivatives. For any fixed number of derivatives, the entropy satisfies a non-perturbative second law of black hole mechanics if the black hole remains within the regime of validity of EFT. In equilibrium the entropy reduces to the Wald entropy. It reduces to the entropy defined by Hollands {\it et al} in theories of vacuum gravity with up to $10$ derivatives.  
\end{abstract}

\maketitle

{\it Introduction.} 
Viewed as an effective field theory (EFT), General Relativity should exhibit higher derivative corrections arising from integrating out massive UV degrees of freedom. If the picture of black holes as thermodynamic objects is correct then the laws of black hole mechanics should remain valid when such corrections are included. For the first law, this was confirmed by Wald \cite{Wald:1993nt,Iyer:1994ys} who defined the entropy of a stationary (time-independent) black hole for any gravitational theory with a diffeomorphism invariant Lagrangian. It is an open problem (except for very special theories \cite{Jacobson:1995uq}) to generalize this to a definition of entropy for a {\it dynamical} black hole. Such a definition should depend only on the local geometry of a ``constant time'' cross-section of the event horizon, satisfy a second law, and reduce to the Wald entropy in equilibrium. 

Recent work on this problem has resorted to perturbation theory around a stationary black hole background. Consider a $1$-parameter family of metrics $g(\epsilon)$ with $g(0)$ a stationary black hole solution. Wall has described an algorithm \cite{Wall:2015} for defining an entropy that reduces to the Wald entropy in equilibrium and satisfies a second law to {\it linear order} in $\epsilon$. (See \cite{Sarkar:2013swa,Bhattacharjee:2015yaa} for special cases. This entropy coincides with expressions for holographic entanglement entropy in vacuum gravity theories with up to $4$ derivatives \cite{Fursaev:2013fta,Dong:2013qoa}.) At linear order, the second law simply asserts that the entropy must be constant. To see an entropy {\it increase}, one must go beyond linear order in $\epsilon$. This has been achieved by Hollands, Kov\'acs and Reall (HKR) \cite{Hollands:2022}, who extended Wall's algorithm to define an entropy that satisfies a second law to {\it quadratic order} in $\epsilon$. This second law holds {\it in the sense of EFT}, which means: (a) It holds for black hole solutions lying within the regime of validity of the EFT, i.e., solutions which vary only on length/time scales large compared to the UV scale; (b) in an EFT defined by terms with up to $N$ derivatives, the second law holds only modulo a correction of the same size as the neglected terms with more than $N$ derivatives.

We shall extend these ideas to obtain a definition of black hole entropy which satisfies a {\it non-perturbative} second law, in the sense of EFT. ``Non-perturbative'' means that, for fixed $N$, it holds to all orders in the amplitude of the perturbation $\epsilon$ (more precisely, it holds without requiring any expansion in $\epsilon$). Similarly to the HKR definition, we define the entropy order by order in derivatives. For 4d vacuum gravity, our definition reduces to that of HKR for theories with $N\le 10$ derivatives (which reduces to \cite{Fursaev:2013fta,Dong:2013qoa,Wall:2015} for $N =4$). So a corollary of our result is that the HKR entropy satisfies a non-perturbative second law in such theories. See \cite{Davies:2022xdq} for explicit expressions for this entropy in some $N=6,8$ theories.

{\it Preliminaries.} Our assumptions are the same as HKR, which we summarize here. 

We shall consider EFTs of vacuum gravity in $d$ dimensions, in which the only field is the metric. We assume that the EFT is specified by a scalar Lagrangian, which is a sum of terms with increasing total number of derivatives $16\pi G{\cal L} = 2\Lambda +  R+ \ldots$. For simplicity we assume that either $d$ is even, or that the theory is invariant under spacetime orientation reversal, which implies that only terms with even numbers of derivatives appear in the Lagrangian \cite{Hollands:2022}. The coefficient of a term on the RHS with $k$ derivatives is assumed to be a dimensionless multiple of $\ell^{k-2}$ where $\ell$ is the UV length scale of the EFT. (Except for $k=0$, see below.)

We consider a $1$-parameter family of black hole solutions of our theory, parameterized by a length scale $L$. For each member in this family we assume the following. (I) The generators of the event horizon $\cH$ are future-complete and the solution ``settles down'' at late time to a stationary solution whose event horizon is a Killing horizon satisfying the zeroth law of black hole mechanics (see \cite{Bhattacharyya:2022nqa,Hollands:2022ajj}). (II) ${\cal H}$ is smooth, with compact cross-sections. Smoothness would be violated at early time in a highly dynamical situation (e.g. a black hole merger) but then the growth in the Bekenstein-Hawking entropy would presumably overwhelm any higher derivative contributions. But smoothness should hold during the ``settling down'' period after a merger (or gravitational collapse), and also when a black hole interacts with weak gravitational waves. (III) The solution lies within the regime of validity of EFT. This means that, near $\cH$, the solution does not vary significantly on length/time scales below $L$, and that $L \gg \ell$. More precisely, in the coordinates introduced below, we have $\partial^k g_{\mu\nu}=O(L^{-k})$ on $\cH$ (for any $k$). We also assume  $\Lambda=O(L^{-2})$, i.e., $\Lambda$ scales as a $2$-derivative quantity (cf the cosmological constant problem). We shall {\it not} perform any expansion of the solution in $L$.

We assume that our EFT is known only up to the terms with $N$ derivatives ($N$ is even). The equation of motion takes the form $E_{\mu\nu} = O(\ell^N/L^{N+2})$ where the LHS comes from the known terms and the RHS is the corrections from the unknown terms with $N+2$ or more derivatives. Henceforth we shall write the RHS simply as $O(\ell^N)$. Powers of $L$ can be reinstated by dimensional analysis. 

{\it Notation.} We mostly follow HKR. By (II) we can use Gaussian null coordinates in a neighbourhood of ${\cal H}$ such that the metric takes the form
$$
 g =-r^2 \alpha dv^2 + 2 dv dr + 2r\beta_A dv dx^A + \mu_{AB}dx^A dx^B
$$
where $A,B=1, \ldots, d-2$ and $\alpha$, $\beta_A$, $\mu_{AB}$ are smooth functions of $(v,r,x^A)$. ${\cal H}$ is at $r=0$ and $v$ is an affine parameter along the horizon generators, with our assumptions valid for $v \ge v_\star$. $\mu_{AB}(v,r,x^A)$ has Riemannian signature and we lower/raise $A,B$ indices with this metric, whose determinant we denote as $\mu$. We define $K_{AB} = (1/2) \partial_v \mu_{AB}$ (describing the expansion and shear of the horizon generators) and $\bar{K}_{AB} = (1/2) \partial_r \mu_{AB}$. A scaling $v' = a^{-1} v$, $r' =a r$ with constant $a$ preserves the form of the metric. A quantity that transforms as $X' = a^b X$ is said to have ``boost weight'' $b$. $(\alpha, \beta_A, \mu_{AB}, K_{AB},\bar{K}_{AB})$ have $b=(0,0,0,1,-1)$. A $v$ ($r$) derivative increases (decreases) $b$ by $1$. $C(v)$ denotes a constant $v$ cross-section of ${\cal H}$. $D_A$ denotes the covariant derivative w.r.t. the metric $\mu_{AB}(v,0,x)$ on this surface. We have the commutator
\be
 \label{commutator}
 [\partial_v, D_A] t_B = (D_C K_{AB}-2 D_{(A}K_{B)C} )t^C
\ee
and similarly for tensors with more indices. All $C(v)$ are diffeomorphic to a single manifold $C$ with coordinates $x^A$. $\NN(v)$ denotes the part of $\cH$ to the future of $C(v)$. Define an integration measure on $\NN(v_0)$ as
$$
    \int_{\NN(v_0)} dV (\ldots) \equiv   \int_{C} d^{d-2} x  \sqrt{\mu(v_0,0,x)} \int_{v_0}^\infty dv (\ldots)
$$
We shall frequently suppress $A,B, \ldots$ indices e.g. $K$ denotes $K_{AB}$, $D^q$ denotes $D_{A_1} \ldots D_{A_q}$ etc. 

{\it The HKR entropy.}
HKR defined an entropy
$$
    S_{\rm HKR}(v) = \int_{C} d^{d-2} x \sqrt{\mu(v,x)} s^{v}_{\rm HKR}(v,x)
$$
where the entropy density $s^{v}_{\rm HKR}$ (the $v$ component of an entropy current) is a linear combination of terms, each with up to $N-2$ derivatives, with each term a product (understood to include possible contraction) of ``allowed terms'', which are: $\mu_{A B}$, $\mu^{A B}$, $D^k \beta_{A}$, $D^k\partial_{v}^p K_{A B}$, $D^k\partial_{r}^q \bar{K}_{A B}$, $\Lambda$, $\epsilon_{A_1 ... A_{d-2}}$ (the volume form on $C(v)$) and $D^k R_{A B C D}[\mu]$ (where $R_{ABCD}$ is the Riemann tensor of $\mu_{A B}$). Note the only allowed terms with $b>0$ are of the form $D^k\partial_{v}^p K_{A B}$. For a stationary black hole, $S_{\rm HKR}$ coincides with the Wald entropy. It also satisfies the first law of black hole mechanics. If assumptions (I,II,III) hold then the HKR entropy satisfies, for any $v_0 \ge v_\star$,
\begin{multline} \label{HKR}
    \dot{S}_{\rm HKR}(v_0) =  \frac{1}{4} \int_{\NN(v_0)} dV \left(||K_{AB} + X_{AB}||^2+   \right. \\
    \left. + D_A Y^A \right)(v,x)   +  O(\ell^N) 
\end{multline} 
where $||T_{AB}||^2 \equiv T_{AB}T^{AB}$. $X_{AB}$ and $Y^A$ are $O(\ell^2)$ with $b=1,2$ respectively. They are linear combinations of products of allowed terms evaluated on $\cH$. Each term in $Y^A$ has at least two factors with $b>0$ and so is a linear combination of terms of the form $(D^{k}{\partial_{v}^{p} K}) \, (D^{k'}{\partial_{v}^{p'} K}) Q$ where $Q$ is a product of allowed terms. Quantities with $b>0$ vanish on the horizon of a stationary black hole satisfying (I) hence, in perturbation theory around such a hole, $Y^A$ is $O(\epsilon^2)$. So, to $O(\epsilon^2)$, the term in \eqref{HKR} involving $Y^A$ can be evaluated in the stationary black hole background, where it drops out because it is a total derivative. The RHS above is then non-negative, modulo $O(\ell^N)$ terms, and so a second law holds to $O(\epsilon^2)$, in the sense of EFT \cite{Hollands:2022}.

This argument does not work beyond $O(\epsilon^2)$ because the integrand is a {\it bilocal} quantity: $dV$ involves $\sqrt{\mu}$ defined at affine time $v_0$, whereas the rest of the integrand is defined at affine time $v$. So $(D_A Y^A)(v,x)$ does not integrate to zero. Integrating this term by parts instead gives
\begin{equation}
\label{Inew}
 \frac{1}{4} \int_{\NN(v_0)} dV \, Y^A(v,x) \partial_A \rho(v_0,v,x)
\end{equation}
where $\rho$ is a biscalar measuring the increase of the area element of $\cH$ from time $v_0$ to time $v$:
$$
 \rho(v_0,v,x) = \log \sqrt{\frac{\mu(v,0,x)}{\mu(v_0,0,x)}}
$$
How might the second law fail beyond $O(\epsilon^2)$? $Y^A$ can contain terms that depend only on {\it derivatives} of $K_{AB}$. So the second law might fail in a situation where $K_{AB}$ is small but its derivatives are large, such that the second term of \eqref{HKR} dominates the first, with a bad sign.

{\it The new idea.} We shall exploit the structure of $X_{AB}$ and $Y^A$ to show that, order by order in derivatives, we can integrate by parts and complete the square on $K_{AB}$ to bring \eqref{HKR} to the form 
$$
\dot{S}(v_0) = \frac{1}{4} \int_{\NN(v_0)} dV ||K_{AB} + Z_{AB}||^2 +  O(\ell^N) 
$$
where on the LHS, $S(v_0)$ may differ (for large enough $N$) from $S_{\rm HKR}(v_0)$ through the addition of an integral over $C(v_0)$ of certain new terms, each a product of allowed terms, each with at most $N-2$ derivatives. On the RHS, $Z_{AB}$ is a {\it bilocal} quantity of boost weight $1$ that is a linear combination of terms, each a product of allowed terms (evaluated at time $v$) but also with possible factors of the form $D^q \rho$ with $q \ge 1$. The RHS has a good sign (modulo higher order EFT corrections) hence {\it $S(v)$ satisfies a non-perturbative second law}, in the sense of EFT. It is non-perturbative because we have not had to perform any expansion in $\epsilon$. The terms modifying $S_{\rm HKR}$ each have at least two factors of positive boost weight and so, if we did expand in $\epsilon$, these terms would be $O(\epsilon^2)$, which implies that $S$ reduces to the Wald entropy in equilibrium, and satisfies the first law. 

{\it The construction.} We shall prove the above result by induction using several results from HKR. We shall describe the proof completely but briefly. A lengthier account, applicable to a larger class of theories, will appear in the companion paper \cite{iain}. 

The inductive hypothesis is that, for even $m$ in the range $0 \le m \le  N-2$ we can define
$$
S_m(v) = S_{\rm HKR}(v) + \int_{C} d^{d-2}x \sqrt{\mu(v)} s_m^v(v)
$$
with $s_m^v$ a linear combination of $b=0$ terms, each with at most $m$ derivatives, each a product of allowed terms, with at least two factors of positive boost weight, such that, for any $v_0\ge v_\star$, 
\begin{multline}
 \label{hypothesis}
    \dot{S}_m(v_0) =O(\ell^N)+\\
    \frac{1}{4} \int_{\NN(v_0)} dV  \left( ||K_{A B} + Z_{mA B}||^2 + R_m \right)
\end{multline}
Here $Z_{m}$ has $b=1$ and is $O(\ell^2)$; $R_m = \sum_n R_{m,n}$ where the sum is over even $n$ in the range $m+2 \le n \le N-2$, $R_{m,n}$ is a linear combination of $(n+2)$-derivative terms of the form $\ell^n (D^{k}{\partial_{v}^{p} K}) \, (D^{k'}{\partial_{v}^{p'} K}) Q$, where $Q$ has boost weight $-p-p'$. Furthermore, $Z_m$ and $Q$ is each a linear combination of terms, where each term is a product of factors of two possible types: (i) allowed terms evaluated at affine time $v$ and (ii) $D^q \rho$ with $q\ge 1$ ($D_A$ evaluated at time $v$). If a factor of type (ii) is present then the term is bilocal, otherwise it is local.

By (\ref{HKR}) and (\ref{Inew}), the base case $m=0$ is satisfied with $s_0^v=0$, $Z_{0A B}= X_{A B}$ and $R_0 = YD \rho$. Now let's assume the hypothesis is true for some even $m<N-2$. The obstruction to it being true for $m+2$ is $R_{m,m+2}$. For each term in $R_{m,m+2}$ we can integrate by parts w.r.t. $x^A$ repeatedly to reduce $k$ to $0$. Each integration by parts either (i) increases $k'$ by $1$; (ii) gives $D$ acting on $Q$, which preserves the assumed form of $Q$; (iii) gives $D$ acting on the measure $dV$: as in \eqref{Inew} this generates an additional factor of $D \rho$, which can be absorbed into $Q$, preserving its assumed form.

Setting $k=0$ leaves terms of the form $\ell^{m+2} (\partial_{v}^{p} K) \, (D^{k'}{\partial_{v}^{p'} K}) Q$. We now aim to reduce $p$ to $0$ by integrating by parts w.r.t. $v$. To avoid surface terms we must treat local and bilocal terms in $Q$ differently. First consider a bilocal term, i.e., one containing at least one factor of $D^q \rho$. Integrating by parts w.r.t. $v$ to reduce $p$ by $1$ generates surface terms at $v=\infty$ and $v=v_0$. The former vanishes because assumption (I) implies that $b>0$ factors vanish at $v=\infty$ \cite{Hollands:2022}. The latter vanishes because of the factor $D^q \rho$, since $\rho=0$ at $v=v_0$. The new integrand is of the form $\partial_v^{p-1} K \partial_v (D^{k'} \partial_v^{p'} K Q)$. Expanding out the $v$-derivative, this can be brought to the form of the inductive hypothesis as follows. (i) use \eqref{commutator} repeatedly in $\partial_v D^{k'} \partial_v^{p'} K$. (ii) Use HKR's result that equations of motion can be used to eliminate non-allowed terms (arising from $\partial_v Q$) such as $\partial_v \beta$ or $\partial_v \bar{K}$ in favour of allowed terms and $O(\ell^2)$ terms; the latter are then absorbed into $R_{m,n}$ for $n \ge m+4$. (iii) Use \eqref{commutator} and $\partial_v \rho \propto K$ to write $\partial_v D^q \rho$ (arising from $\partial_v Q$) in terms of $D^qK$ and $D^r K D^{q-r} \rho$ ($r=1,\ldots, q-1$; note $[\partial_v,D]\rho=0$). After these steps we have a sum of terms of the same form as the start of this paragraph but with $p$ reduced by $1$. However, some of these terms might be local (unlike our original term) because in step (iii) a $v$-derivative can convert bilocal $D^q \rho$ into local $D^q K$. 
By repeating this procedure on the new set of bilocal terms we can eventually reduce our original term to a sum of terms, each of which either has $k=p=0$ or has a factor of the form $D^{q}K$ for some $q \ge 1$. In the latter terms we can integrate by parts w.r.t. $x^A$ as in the previous paragraph to reduce $q$ to $0$, giving a term proportional to $K$, i.e., one with $k=p=0$. Thus we have transformed our original bilocal term in $Q$ into a sum of terms with $k=p=0$.

Next we consider local terms with $k=0$. $Q$ is a sum of terms, each a product of allowed terms. In this case, HKR proved (section 4.1) that one can write $\partial_v^p K D^{k'} \partial_v^{p'}K Q$ as the sum of (i) $\partial_{v}\left\{\frac{1}{\sqrt{\mu}} \partial_{v}\left[\sqrt{\mu} \zeta_m^{v} \right] \right\}$; (ii) terms of the form $K \, (D^{\bar{k}}{\partial_{v}^{\bar{p}'} K}) \, \tilde{Q}_1$; (iii) terms of the form $(\partial_{v}^{\bar{p}} K) \, (D^{\bar{k}} K ) \, \tilde{Q}_2$; where $\zeta_m^v$ and $\tilde{Q}_i$ are constructed only from allowed terms (and so are local), with each term in $\zeta_m^v$ containing at most $m+2$ derivatives and at least two factors of positive boost weight. Terms of type (ii) are proportional to $K$ and hence have $k=p=0$. Terms of type (iii) can be integrated by parts w.r.t. $x^A$ to reduce $\bar{k}$ to zero, as described above, then these terms also have $k=p=0$ (this may introduce $D^q \rho$ factors but that does not matter). For terms of type (i), the $v$-integration can be performed explicitly and these terms moved to the LHS of \eqref{hypothesis}, by defining $s^v_{m+2} = s^v_m + \zeta_m^v$. 

After these steps, the terms from $R_{m,m+2}$ remaining on the RHS all have $k=p=0$, i.e., they have a factor of $K$. Hence in \eqref{hypothesis} we can complete the square on $K$ to eliminate these terms. This gives $Z_{m+2}=Z_{m}+O(\ell^{m+2})$. Since $Z_m$ is $O(\ell^2)$, the ``error terms'' resulting from completing the square are $O(\ell^{m+4})$ and of the same form described at the start of the induction. Thus we have shown that \eqref{hypothesis} holds also for $m+2$, completing the induction. Our main result then follows by setting $m=N-2$ with $Z_{AB} = Z_{(N-2)AB}$. 

{\it Difference from HKR.} Our entropy differs from the HKR entropy because of the terms $\zeta_m^v$, arising from a $v$-integration by parts of a {\it local} term in $R_m$. For what $m$ is this first required? Initially, for $m=0$, all terms in $R_0=YD\rho$ are bilocal. Local terms are generated by integration by parts w.r.t. $v$. This is first required when a term in $R_m$ contains a $5$-derivative factor $(\partial_v K) \, (\partial_v K) D \rho$. However, $R_m$ has $b=2$ so there must also be a factor of $\bar{K}^2$ or $\partial_r \bar{K}$, which adds $2$ more derivatives. All terms must have even numbers of derivatives so there must be an extra $D$ somewhere hence $m \ge 4$ e.g. consider a term in $R_{4,6}$ of the form $(\partial_v K) \, (\partial_v K) \, \bar{K} \, (D\bar{K}) \, D \rho$ (other terms behave similarly). When we perform the $v$ integration by parts, we get a local term when $\partial_v$ hits $D \rho$, producing $K \, (\partial_v K) \, \bar{K} \, (D\bar{K}) \, D K$. This is still fine, because it is proportional to $K$ and so can be absorbed into the square, generating a $10$-derivative ``error term'' in $R_{6,8}$. But this error term is proportional to $DK$, and so integration by parts w.r.t. $x^A$ renders it proportional to $K$. Therefore, it too can be absorbed by completing the square at $m=6$. It is only in $R_{8,10}$ (12 derivatives) that we might first need to perform a $v$ integration by parts on a local term. Thus $\zeta_m^v=0$ for $m \le 6$ so $s_m^v=0$ for $m \le 8$, i.e., our entropy agrees with the HKR entropy for $N \le 10$. 

{\it Uniqueness.} Our construction demonstrates existence of an entropy satifying a second law, but not its uniqueness. It seems plausible that, beyond linear perturbations around equilibrium, there is not a unique definition of black hole entropy \cite{Hollands:2022}. 

{\it Gauge invariance.} The HKR entropy is not manifestly invariant under a change of Gaussian null coordinates \cite{Hollands:2022}, so it is not determined just by the geometry of a horizon cross-section: a choice of gauge is also required. The same is true of our entropy. However, gauge non-invariant terms do not occur for $N \le 6$ \cite{Davies:2022xdq}. So, for EFTs with $N \le 6$ derivatives, the HKR entropy is gauge invariant and, by our result, satisfies a non-perturbative second law. For $N=8$, gauge non-invariant terms do arise \cite{Davies:2022xdq}. The implications of a lack of gauge-invariance are discussed in section 1.11 of \cite{Hollands:2022}.

{\it Einstein-Maxwell-scalar EFT.} In the companion paper \cite{iain} it is shown that our definition of black hole entropy generalizes straightforwardly to EFTs of gravity coupled to a Maxwell field and scalar field and that, in such theories, the entropy is again gauge-invariant for $N \le 6$.

{\it Acknowledgments.} We thank S. Hollands and \'A. Kov\'acs for reading a draft and S. Bhattacharyya for helpful discussions.  ID is supported by an STFC studentship. HSR is supported by STFC grant no. ST/X000664/1.

\end{document}